\documentclass[prb,showpacs,preprintnumbers,amsmath,amssymb,printfigures,twocolumn,superscriptaddress]{revtex4-1}
\usepackage[colorlinks]{hyperref}
\usepackage{graphicx}
\usepackage{amsmath}
\usepackage{bm}

\usepackage[T1]{fontenc}
\usepackage[latin9]{inputenc}
\usepackage{textcomp}
\usepackage{amstext}
\usepackage[Symbol]{upgreek}
\usepackage{graphicx}
\usepackage{amsmath}
\usepackage{booktabs}
\usepackage{hyperref}
\usepackage{dcolumn}
\usepackage{color}
\usepackage{enumerate}
\usepackage{latexsym,bm}
\makeatletter

\begin{document}
\preprint{manuscript}

\title{Native point defects in few-layer phosphorene}

\author{V. Wang}
\thanks{wangvei@icloud.com (V. Wang).}
\affiliation{Department of Applied Physics, Xi'an University of Technology, Xi'an 710054, China}  

\author{Y. Kawazoe}
\affiliation{New Industry Creation Hatchery Center, Tohoku University, Sendai, Miyagi 980-8579, Japan} 
\affiliation{Kutateladze Institute of Thermophysics, Siberian Branch of Russian Academy of Sciences, Novosibirsk 630090, Russia}

\author{W.T. Geng}
\thanks{geng@ustb.edu.cn (W.T. Geng).}
\affiliation{School of Materials Science \& Engineering, University of Science and Technology Beijing, Beijing 100083, China}
\affiliation{Psi Quantum Materials LLC, Laiwu 271100, China}
\date{\today}

\begin{abstract}
Using hybrid density functional theory combined with a semiempirical van der Waals dispersion correction, we have investigated the structural and electronic properties of vacancies and self-interstitials in defective few-layer phosphorene. We find that both a vacancy and a self-interstitial defect are more stable in the outer layer than in the inner layer. The formation energy and transition energy of both a vacancy and a self-interstitial P defect decrease with increasing film thickness, mainly due to the upward shift of the host valence band maximum in reference to the vacuum level. Consequently, both vacancies and self-interstitials could act as shallow acceptors, and this well explains the experimentally observed 
\emph{p}-type conductivity in few-layer phosphorene. On the other hand, since these native point defects have moderate formation energies and are stable in negatively charged states, they could also serve as electron compensating centers in \emph{n}-type few-layer phosphorene.
\end{abstract}
\maketitle 

\section{introduction}

The successful fabrication of two-dimensional materials such as graphene and transition metal dichalcogenides arouses intense interest of researchers with intriguing electronic, mechanical, optical, and thermal properties.\cite{Novoselov2005, Zhang2005, Radisavljevic2011, Wang2012} The gapless nature of graphene and low carrier mobility of transition metal dichalcogenides, however, present limitations to their potential application in  industry.\cite{Liao2010, Schwierz2010, Wu2011, Mak2010, Radisavljevic2011} Very recently, another exciting two-dimensional material, few-layer black phosphorus name as phosphorene, has been successfully fabricated.\cite{Li2014, Dai2014, Reich2014, Liu2014} The phosphorene-based field effect transistor exhibits a carrier mobility up to 1000 cm$^{2}$/V$\cdot$s and an of/off ration up to 10$^4\sim10^5$.\cite{Li2014,Liu2014,Koenig2014} 

Similar to graphite, black phosphorus is also a layered material held together by interlayer van der Waals (vdW) interactions. Inside a layer, each phosphorus atom bonds with three nearest neighbors by sharing all three valence electrons for \emph{sp}$^3$ hybridization in a puckered honeycomb structure.\cite{Rodin2014} Black phosphorus has a direct band gap of 0.31$\sim$0.35 eV.\cite{Keyes1953, Maruyama1981, Akahama1983, Warschauer2004} The band gap of phosphorene has been found to depend on the film thickness. First-principles calculations demonstrated that the energy band gap decreases from 1.5 $\sim$2.0 eV for a monolayer to $\sim$0.6 eV for a five-layer phosphorene.\cite{Qiao2014, Tran2014}  It was also predicted that under strain, few-layer phosphorene could go through a semiconductor-to-metal or direct-to-indirect band gap transition.\cite{Peng2014, Rodin2014} Most recently, Liu \emph{et al.} constructed an inverter using MoS$_2$ as an \emph{n}-type transistor and phosphorene as a \emph{p}-type transistor, and integrated the two on the same Si/SiO$_2$ substrate.\cite{Liu2014} They observed unintentional \emph{p}-type conductivity with high hole mobility in few-layer phosphorene. Additionally, a number of experiments have also achieved intrinsic \emph{p}-type phosphorene.\cite{Yuan2014, Li2014, Liu2014, Deng2014, Das2014} 

Then a question arises: what is the origin of the reported intrinsic \emph{p}-type conductivity in phosphorene? 
Defects and impurities are usually unavoidable in real materials and often change dramatically the electrical, optical and magnetic properties of three- \cite{Tilley2008} and two-dimensional semiconductors.\cite{Terrones2012, Tongay2013, Qiu2013, Zhu2014}  A large number of theoretical studies on the thickness-dependence of the electronic structure of few-layer phosphorene notwithstanding, knowledge of the properties of native point defects in few-layer phosphorene is still missing. In the present work, we have investigated the formation energies and transition levels of both vacancies and self-interstitials in few-layer phosphorene by performing first-principles calculations using hybrid density functional \cite{Becke1993, Perdew1996a, Heyd2003} in combination with a semiempirical vdW correction approach developed by Grimme and co-workers\cite{Grimme2006}, aiming  to elucidate the origin of unintentional \emph{p}-type conductivity displayed by this novel material. Our calculations demonstrated that: (\emph{i}) the host band gap, formation energies and acceptor transition levels of both vacancies and self-interstitials all decrease with increasing film thickness of phosphorene; (\emph{ii}) both native point defects are possible sources of the intrinsic \emph{p}-type conductivity manifested in few-layer phosphorene; (\emph{iii}) these native defects have low formation energies and thus could serve as compensating centers in \emph{n}-type multilayer phosphorene. The remainder of this paper is organized as follows. In Sec. II, methodology and computational details are described. Sec. III presents the calculations of formation energies and transition energies of native point defects in few-layer phosphorene, followed by electronic structure analyses. Finally, a short summary is given in Sec. IV.

\section{Methodology}
Our total energy and electronic structure calculations were carried out using the VASP code, \cite{Kresse1996, Kresse1996a}, based on the hybrid density functional theory (DFT) proposed by Heyd, Scuseria, and Ernzerhof (HSE).\cite{Heyd2003} The recent development of hybrid DFT can yield band gaps in good agreement with measurements,\cite{Paier2006, Marsman2008, Park2011} and thus provide more reliable description of transition levels and formation energies of defects in semiconductors.\cite{Alkauskas2011, Deak2010, Komsa2011, Freysoldt2014} We here have employed a revised scheme, HSE06.\cite{Krukau2006} The screening parameter was set to 0.2 {\AA}$^{-1}$; the Hartree-Fock (HF) mixing parameter $\alpha$ was tuned to produce a band gap similar to the one given by the GW0 approximation, \cite{Hedin1965, Shishkin2006} which means that $\alpha$\% of HF exchange with (100-$\alpha$)\% of Perdew, Burke and Ernzerhof (PBE) exchange\cite{Perdew1996} in the generalized gradient approximation (GGA) were mixed and adopted in exchange functional. The core-valence interaction was described by the frozen-core projector augmented wave (PAW) method.\cite{PAW, Kresse1999} The electronic wave functions were expanded in a plane-wave basis with a cutoff of 250 eV. Test calculations show that the calculated formation energies of neutrally and negatively charged P vacancy in monolayer phosphorene will change by less than 0.1 eV if the energy cutoff is increased to 400 eV. Previous theoretical calculations have shown that the interlayer vdW interaction need to be considered  for a proper description of the geometrical properties of black phosphorus.\cite{Appalakondaiah2012} We therefore incorporated the vdW interactions by employing a semiempirical correction scheme of Grimme's DFT-D2 method, which has been successful in describing the geometries of various layered materials.\cite{Grimme2006, Bucko2010}

\begin{figure}[htbp]
\centering
\includegraphics[scale=0.5]{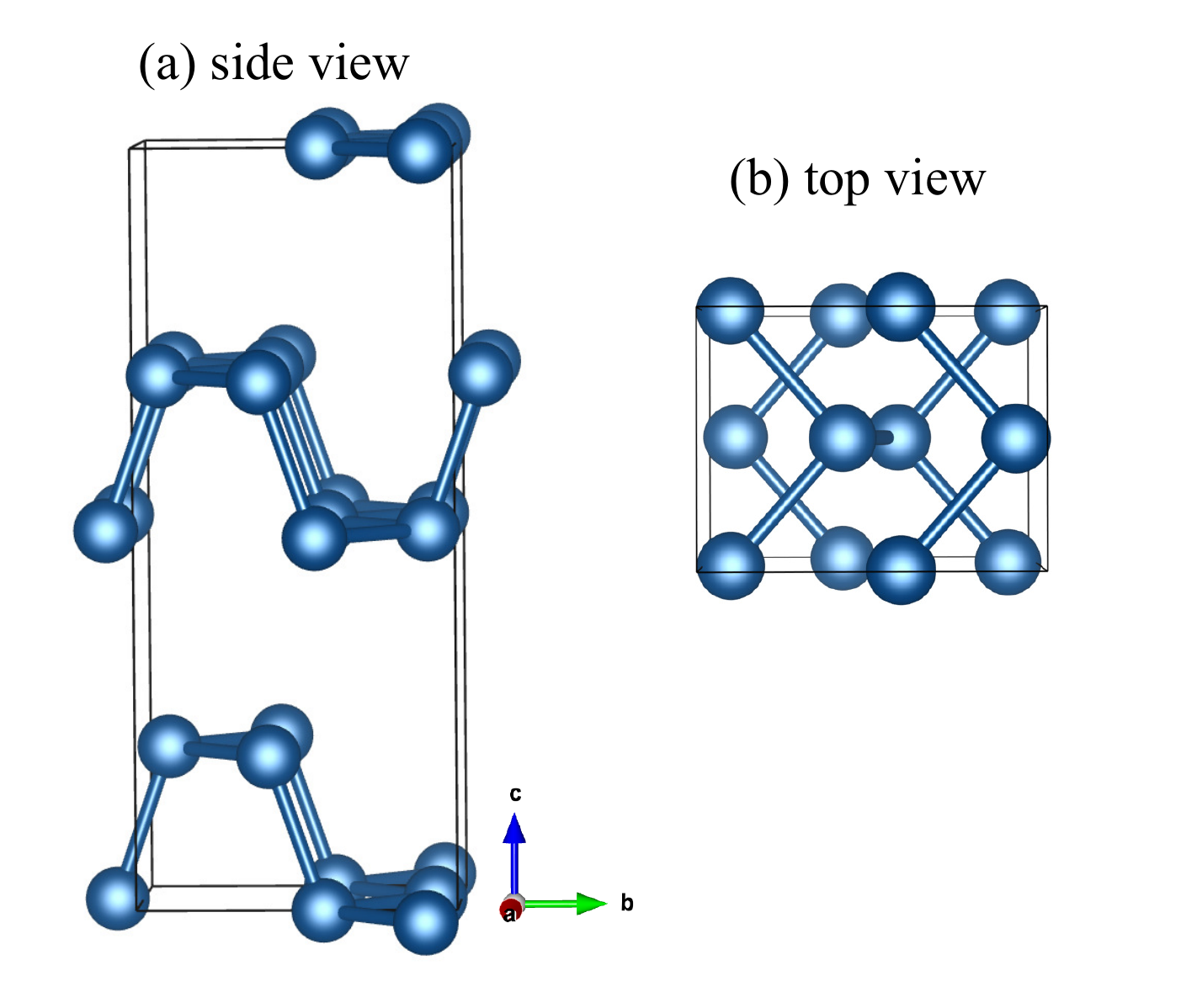}
\caption{\label{bulk}(Color online) Top (a) and side (b) views of the unit cell of black phosphorus.}
\end{figure}

In simulation, a thin film of black phosphorus can be easily obtained by simply truncating the bulk into a slab containing only a few atomic layers. The atomic structure of the black phosphorus is presented in Fig. \ref{bulk}, from which a layered structure is clearly seen. In each layer, the \emph{sp}$^3$  hybridization between one P atom and its three neighbors lead to the tripod-like local structure along \emph{c} direction. In the slab model of few-layer phosphorene, periodic slabs were separated by a vacuum no thinner than 15 {\AA}. For bulk black phosphorus, an 8\texttimes{}6\texttimes{}1 \emph{k}-mesh including $\Gamma$-point, generated according to the Monkhorst-Pack scheme,\cite{Monkhorst1976} was applied to the Brillouin-zone integrations. On geometry optimization, both the shapes and internal structural parameters of pristine unit-cells were fully relaxed until the residual force on each atom less than 0.01 eV/{\AA}. 

\begin{figure}[htbp]
\centering
\includegraphics[scale=0.40]{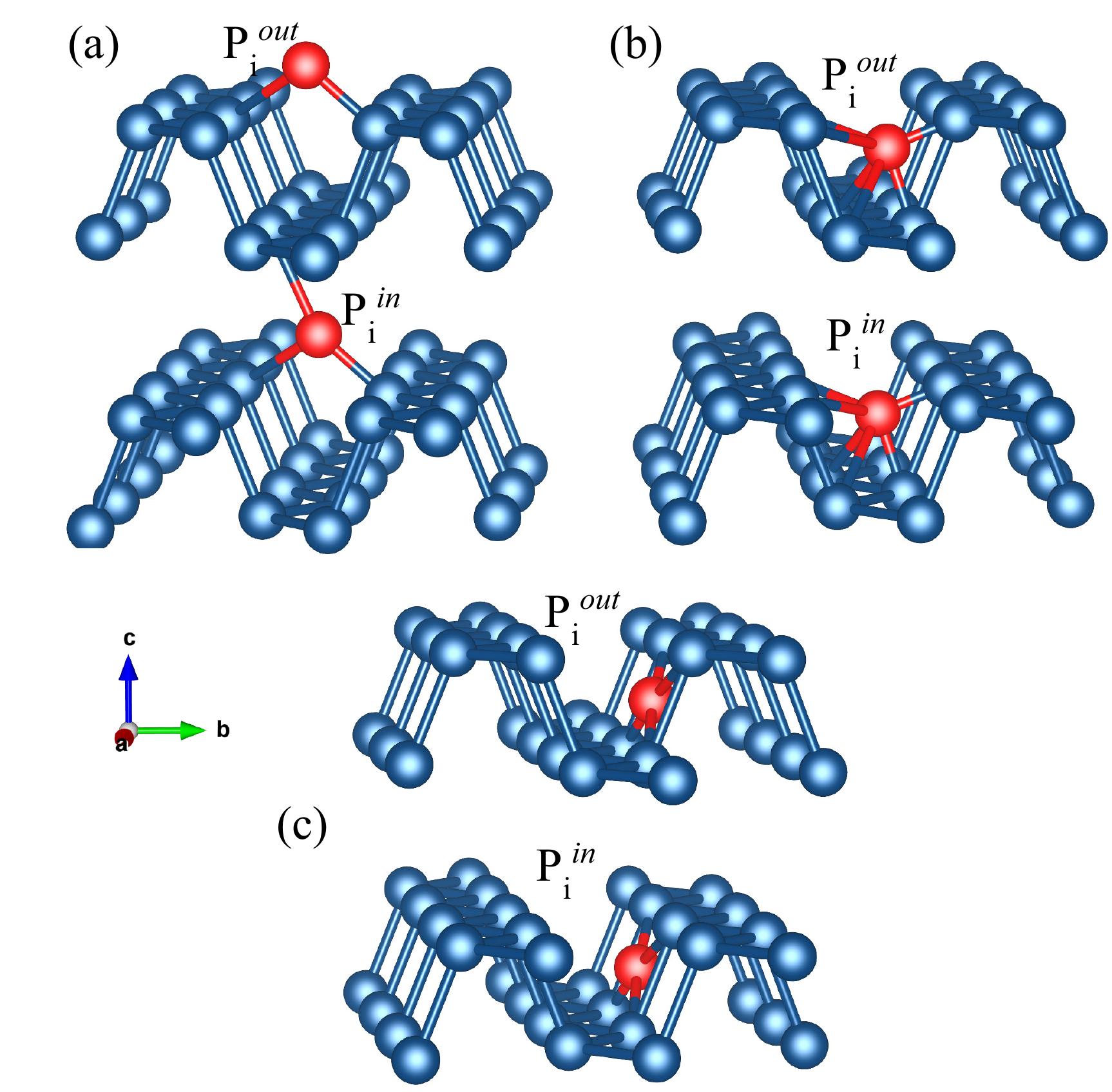}
\caption{\label{interstitial}(Color online) Six inequivalent interstitial configurations in phosphorene bilayer. The point defects are colored differently.}
\end{figure}

The defective system containing a self-interstitial atom, P$_i$, or a vacancy, V$_\text{P}$, was modeled by adding or removing a P atom to or from a 3\texttimes{}2 supercell of few-layer phosphorene. They were the native point defects considered in the present work. In a monolayer phosphorene, there are three interstitial sites; whereas in a multi-layer film, both P$_i$ and V$_\text{P}$ can reside either in the outer or inner layers. We label these positions as X$^{in}$ and X$^{out}$ (X=P$_i$ and V$_\text{P}$) respectively. In Fig. \ref{interstitial}, we show the six inequivalent interstitial sites in a bilayer phosphorene. In view of the fact that the contribution of vdW interaction to the stability of adsorbate on graphene, even in the chemisorption cases, is non-negligible, \cite{Wang2012b} we expected that the HSE06 plus DFT-D2 method should give a more accurate description on the local structure of interstitial defects in few-layer phosphorene. A $\Gamma$-centered 2\texttimes{}2\texttimes{}1 Monkhorst-Pack \emph{k}-mesh was adopted for the 3\texttimes{}2\texttimes{}1 supercells. The internal coordinates in the defective supercells were relaxed to reduce the residual force on each atom to less than 0.02 eV/{\AA}. Moreover, we have allowed spin-polarization for defective systems. A more detailed discussion on the convergence of total energies of defective systems with respect to vacuum thickness is given in the next section.

An accurate description of the band structure of phosphorene is a prerequisite for obtaining reliable predictions on defect properties, which impact greatly the electronic conductivity in phosphorene. Since there is no reported experimental data for the band gaps of few-layer phosphorene, we compare our HSE06 results for defect-free few-layer phosphorene with those obtained using highly accurate quasiparticle GW0 calculations \cite{Hedin1965, Shishkin2006}. The GW0 approximation has been shown to provide very reliable descriptions of the electronic and dielectric properties for many semiconductors and insulators.\cite{Fuchs2007, Shishkin2007} To achieve good convergence of dielectric function in the GW0 calculations, we used a large number of energy band, 80 times of the total number of involved atoms. The converged eigenvalues and wavefunctions obtained from HSE06 with 25\% HF exact exchange (denoted as HSE06-25\% hereafter) functional were chosen as the initial input for the GW0 calculations. 
Note that in GW0 calculations only the quasiparticle energies were recalculated self-consistently in four iterations;
the wavefunctions were not updated but remain fixed at the HSE06-25\% level. A 200 frequency grid points was applied to the integration over the frequencies along the imaginary time axis and real axis. For visualization purpose, the GW0 bands were interpolated based on Wannier orbitals, implemented in WANNIER90 code.\cite{Mostofi2008}

To model a charged-defect, a uniform background charge with opposite sign was added to keep the global charge neutrality of the whole system. The formation energy of a charged defect was defined as  \cite{Zhang1991}

\begin{equation}\label{eq1}
\begin{split}
\Delta E^f_D(\alpha,q)=E_{tot}(\alpha,q)-E_{tot}(host,0)-n_{\alpha}\mu_{\alpha}  \\ 
+q(\mu_{e}+\epsilon_{v})+E_{corr}[q],
\end{split}
\end{equation}

where $E_{tot}(\alpha,q)$ and $E_{tot}(host,0)$ are the total energies of the supercells with and without defect $\alpha$. \emph{n}$_\alpha$ is the number of atoms of species $\alpha$ added to (\emph{n}$_\alpha$>0) or and removed from (\emph{n}$_\alpha$<0) the perfect supercell to create defect $\alpha$. $\mu_{\alpha}$ is the atomic chemical potential equal to the total energy per atom in its elemental crystal.\emph{q} is the charge state of defect, $\epsilon_{v}$ is the host valence band maximum (VBM) level and $\mu_{e}$ is electron chemical potential in reference to the $\epsilon_{v}$ level. Therefore, $\mu_{e}$ can vary between zero and the band-gap (\emph{E}$_g$) of few-layer phosphorene. The final term accounts for both the alignment of the electrostatic potential between the bulk  
and defective (charged) supercells, as well as the finite-size effects resulting from the long-range Coulomb interaction of charged defects in a homogeneous neutralizing background. It can be evaluated by using the Freysoldt correction scheme with an average static dielectric constant.\cite{Freysoldt2009} 

A 12\texttimes{}8\texttimes{}1 \emph{k}-mesh with a Gaussian smearing of 0.01 eV was employed in the calculations of static dielectric tensors $\epsilon$ of pristine few-layer phosphorene. For the static dielectric tensors, the ion-clamped contribution was calculated from the response theory of insulators in finite electric field.\cite{Souza2002} Since the ionic contributions depend on the Born-effective charges and the vibrational modes only,\cite{Paier2009} they were treated using GGA-PBE approximation based on density-functional perturbation theory.\cite{Wu2005} More details of this technique can be found in our previous work.\cite{Wang2014} 
The defect thermodynamic transition (ionization) energy level $\epsilon_{\alpha}$(\emph{q}/$\emph{q}^{\prime}$) is defined as the Fermi-level (\emph{E}$_\text{F}$) position for which the formation energies of these charge states are equal for the same defect,
\begin{equation}\label{eq3}
\epsilon_{\alpha}(q/q^{\prime})=[\Delta E^f_D(\alpha,q)-\Delta E^f_D(\alpha,q^{\prime})]/(q^{\prime}-q).
\end{equation}
More specifically, the defect is stable in the charge state \emph{q} when the \emph{E}$_\text{F}$ is below $\epsilon_{\alpha}(q/q^{\prime})$, while the defect is stable in the charge state q$^{\prime}$ for the \emph{E}$_\text{F}$ positions above $\epsilon_{\alpha}(q/q^{\prime})$.

\section{Results and discussion}
\subsection{Fundamental properties of pristine few-layer phosphorene}
Prior to the investigation of defective system, we have first calculated the atomic and electronic properties of pristine few-layer phosphorene. The calculated lattice parameters as a function of film thickness, yielded by PBE, PBE+vdW, and HSE06-25\%+vdW treatments of the density functional are listed in Table \ref{lattice}. 
We find the lattice parameter \emph{b} increases by 0.07-0.15 {\AA} from bulk to monolayer, while \emph{a} and interlayer distance $\Delta$\emph{d} are quite insensitive to the film thickness. Similar trends were also reported in a previous first-principles study by Qiao \emph{et al.}\cite{Qiao2014} For bulk black phosphorus, the measured lattice parameters are \emph{a}=3.31 {\AA}, \emph{b}=4.38 {\AA} and $\Delta$\emph{d}=5.24 {\AA}.\cite{Brown1965} We see that PBE overestimates both \emph{b} (3.6\%) and $\Delta$\emph{d} (5.5\%); PBE+vdW and HSE06+vdW, on the other hand, are in much better agreement with experiment. So far, there are no experimental data for few-layer phosphorene systems, but we speculate that the success of PBE+vdW and HSE06+vdW in description of bulk black phosphorus could probably extend to few-layer phosphorene. Therefore, we include vdW correction in the following calculations unless otherwise stated.

\begin{table*}[htbp]
\centering
\begin{ruledtabular}
\caption{\label{lattice} Lattice constants \emph{a}, \emph{b} and interlayer distance $\Delta$\emph{d} as a function of film thickness in few-layer phosphorene given by PBE, PBE+vdW and HSE06-25\%+vdW approaches respectively.}
\begin{tabular}{c|ccc|ccc|ccc|}
&\multicolumn{3}{c|}{PBE}
&\multicolumn{3}{c|}{PBE+vdW}
&\multicolumn{3}{c|}{HSE06-25\%+vdW}\\
Systems & \emph{a} (\AA) & \emph{b} (\AA) &  $\Delta$\emph{d} (\AA) & \emph{a} (\AA) & \emph{b} (\AA)  &  $\Delta$\emph{d} (\AA) & \emph{a} (\AA)& \emph{b} (\AA) &  $\Delta$\emph{d} (\AA) \\
\hline
monolayer &3.30 & 4.61 & - &   3.32 &  4.56  &  - &  3.30 &  4.50 & - \\
bilayer   & 3.31 & 4.58 & 5.57 & 3.32 &  4.50  & 5.21	& 3.30 &   4.45 & 5.17  \\
trilayer  & 3.31 & 4.58 & 5.58 & 3.32 &  4.47  & 5.22	& 3.30 &   4.44 & 5.18  \\
quadrilayer & 3.31 & 4.57& 5.59 & 3.32 & 4.46  & 5.23 & 3.30  & 4.44 & 5.19 \\
bulk$^a$ &3.31 & 4.54 & 5.53 & 3.33 & 4.41 & 5.23  &   3.31 &  4.37 & 5.19 \\
\end{tabular}
\leftline{$^a$ Experimental lattice conctants: \emph{a}=3.31 {\AA}, \emph{b}=4.38 {\AA} and $\Delta$\emph{d}=5.24 {\AA} in reference \onlinecite{Brown1965}.}
\end{ruledtabular}
\end{table*}

The standard HSE06 approach with 25\% exact exchange is known to well reproduce the band gaps of small- to medium-gap systems, but not those of wide-gap materials.\cite{Paier2006, Paier2006a, Marsman2008} Recently, Fuchs \emph{et al.} have shown that GW0 approach can describe very well (but slightly overestimate) the electronic structure of wide-gap materials, and the mean absolute relative error (MARE) on the calculated band gaps of some representative traditional semiconductors is only 8.0\%.\cite{Fuchs2007} We summarize the PBE, HSE06, and GW0 calculated band gap of few-layer phosphorene and bulk phosphorus in Table \ref{bandgap}. For the bulk, GW0 gives a band gap of 0.65 eV, significantly higher than the experimental value of 0.31-0.35 eV.\cite{Keyes1953, Maruyama1981, Akahama1983, Warschauer2004} The HSE06 result, 0.28 eV, is slightly lower than experimental value. We therefore expect that GW0 and HSE06-25\% approaches would give a reasonable upper and lower bounds for the band gap of few-layer phosphorene.

The most important knowledge learn from Table \ref{bandgap} is that all density functional forms predict a similar trend: the energy band gap of phosphorene decreases with increasing film thickness. This phenomenon, we argue, is mainly due to the energy band broadening induced by interlayer interaction. Additionally, the quantum confinement effect in low dimensional materials are likely to contribute to this trend.\cite{Kang2013} Since there is no experimental results concerning defective phosphorene and the GW0 approach can perform neither structural optimization nor total energy calculations, we chose to utilize somewhat larger Hartree-Fock mixing parameters $\alpha_\text{opt}$ for thin phosphorene, \emph{i.e.}, 35\% for monolayer and 30\% for bilayer, in an attempt to rectify the probably underestimated band gaps. As for the quadrilayer phosphorene, we used a parameter of 25\%, the same value as for the bulk.

\begin{table*}[htbp]
\centering
\begin{ruledtabular}
\caption{\label{bandgap} The calculated band gap (\emph{E}$_g$) of few-layer phosphorene as a function of film thickness using PBE, HSE06 and GW0 methods respectively.}
\begin{tabular}{c|cccccc}
Systems & PBE & HSE06-25\% & HSE06-$\alpha_\text{opt}$  &  GW0 & Previous work$^a$ & Exp. \\
\hline
monolayer &0.91   &  1.56 &  1.91$^b$ & 2.41 & 1.5-2.0 & - \\
bilayer &0.45  & 1.04 &  	1.23$^c$ & 1.66 & 1.0-1.3 &- \\
trilayer &0.20  & 0.74 &  	0.98$^c$ & 1.20 & 0.7-1.1 &- \\
quadrilayer & 0.16 & 0.71 & 0.71$^d$   & 1.08 &0.5-0.7 & -\\
bulk  &0.10 & 0.28 &  0.28$^d$  & 0.58  & $\sim$0.3  & 0.31$\sim$0.35$^e$ \\
\end{tabular}
\leftline{$^a$ References \onlinecite{Rodin2014,Tran2014,Qiao2014}.}
\leftline{$^b$ HSE06-35\%.}
\leftline{$^c$ HSE06-30\%.}
\leftline{$^d$ HSE06-25\%.}
\leftline{$^e$ References \onlinecite{Keyes1953,Maruyama1981,Akahama1983,Warschauer2004}.}
\end{ruledtabular}
\end{table*}

\begin{figure}[htbp]
\centering
\includegraphics[scale=0.5]{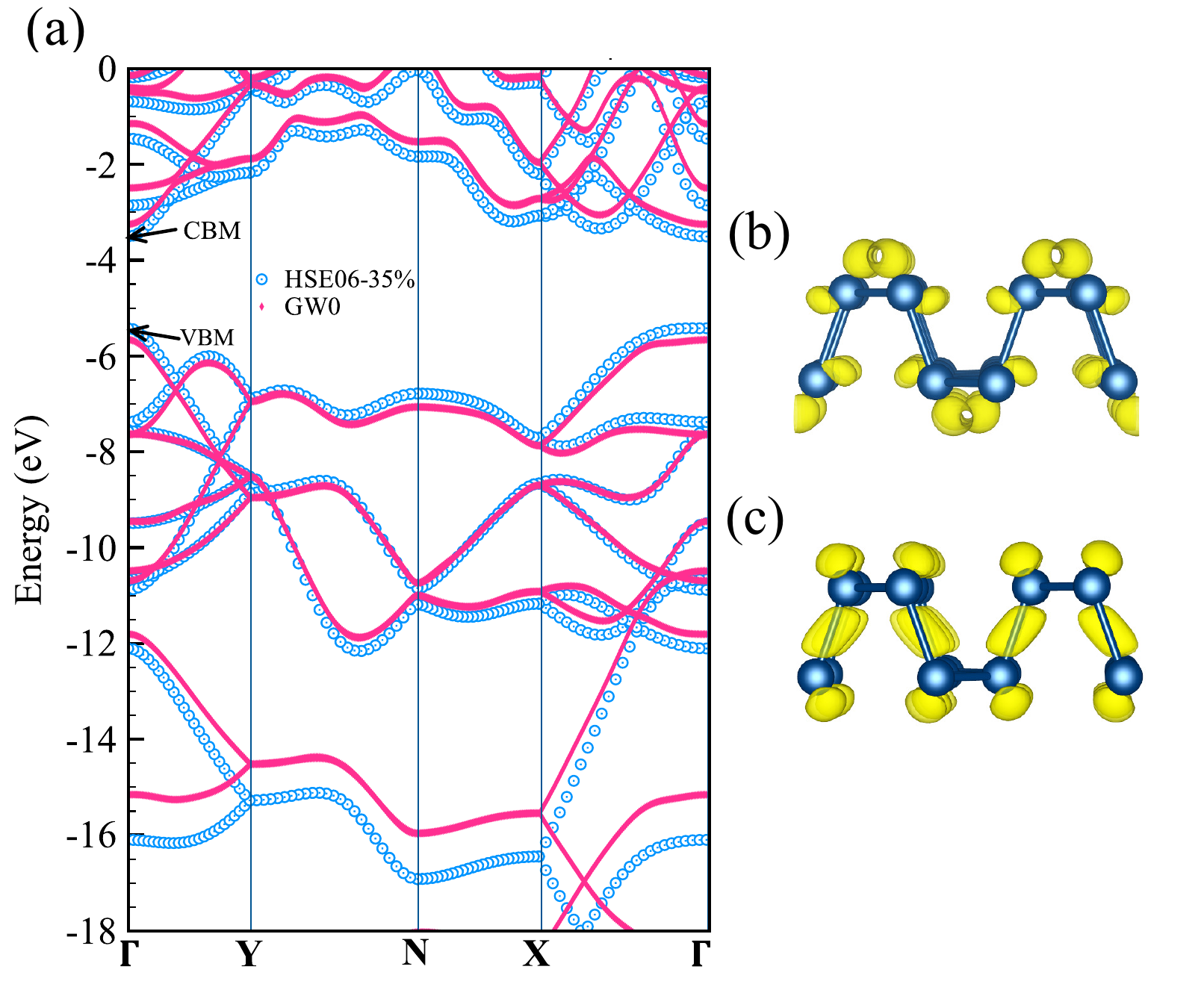}
\caption{\label{monolayer}(Color online) (a) (Color online) Energy band structures (a) of phosphorene monolayer calculated using HSE06 and GW0 methods, and side views of charge density of (b) CBM and (c) VBM. The vacuum level is set to zero and the charge density isosurface levels are shown at 40\% of their maximum values.}
\end{figure}

Figure \ref{monolayer} displays the calculated band structure of monolayer phosphorene using HSE06 and GW0. Note that both the VBM and conduction band minimum (CBM) are located at $\Gamma$-point, and hence a direct band gap. This result is consistent with many previous theoretical studies.\cite{Qiao2014, Tran2014, Liu2014, Peng2014, Rodin2014} However, there is a disagreement on this point. For example, Li \emph{et al.} have argued that monolayer phosphorene possibly possesses an indirect
band gap, because the band interactions near the $\gamma$ point are complicated, as was viewed from a $k\cdot p$ perturbation theory.\cite{Li2014a} The partial charge density analyses show that the VBM are derived from the bonding states between P atoms in different sublayers and the anti-boding states between P atoms in the same sublayer. The opposite is true for the case of CBM. 

\begin{figure}[htbp]
\centering
\includegraphics[scale=0.52]{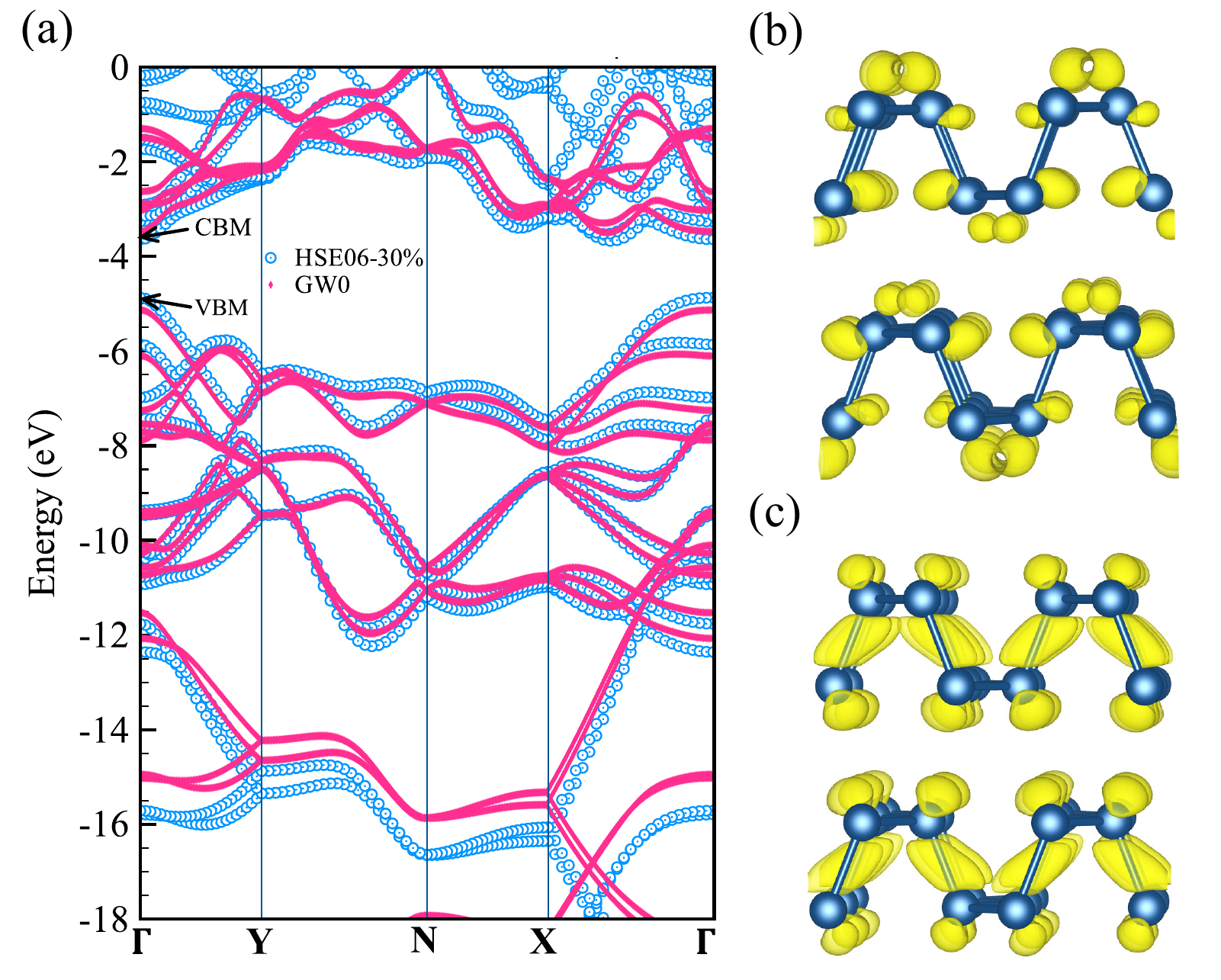}
\caption{\label{bilayer}(Color online) Energy band structures (a) of phosphorene monolayer calculated using HSE06 and GW0 methods, and side views of charge density of (b) CBM and (c) VBM. The vacuum level is set to zero and the charge density isosurface levels are shown at 40\% of their maximum values.}
\end{figure}

We plot in Fig. \ref{bilayer}(a) the band structure of bilayer phosphorene. Clearly, the band characteristics are similar to those of the monolayer, except that in the bilayer, energy level splitting occurs due to the interlayer interactions. The formation of a bilayer phosphorene can be viewed as the result of two monolayer moving close to each other. The degenerated energy levels of two monolayers become non-degenerated via interlayer interactions. Overall, in both monolayer and bilayer cases, HSE06 and GW0 yield similar band dispersion. Remarkable discrepancy occurs to valence states lying 10 eV below the VBM. Energy bands calculated using HSE06 approach are pushed further downward compared to those obtained using GW0 approach.

\begin{figure}[htbp]
\centering
\includegraphics[scale=0.48]{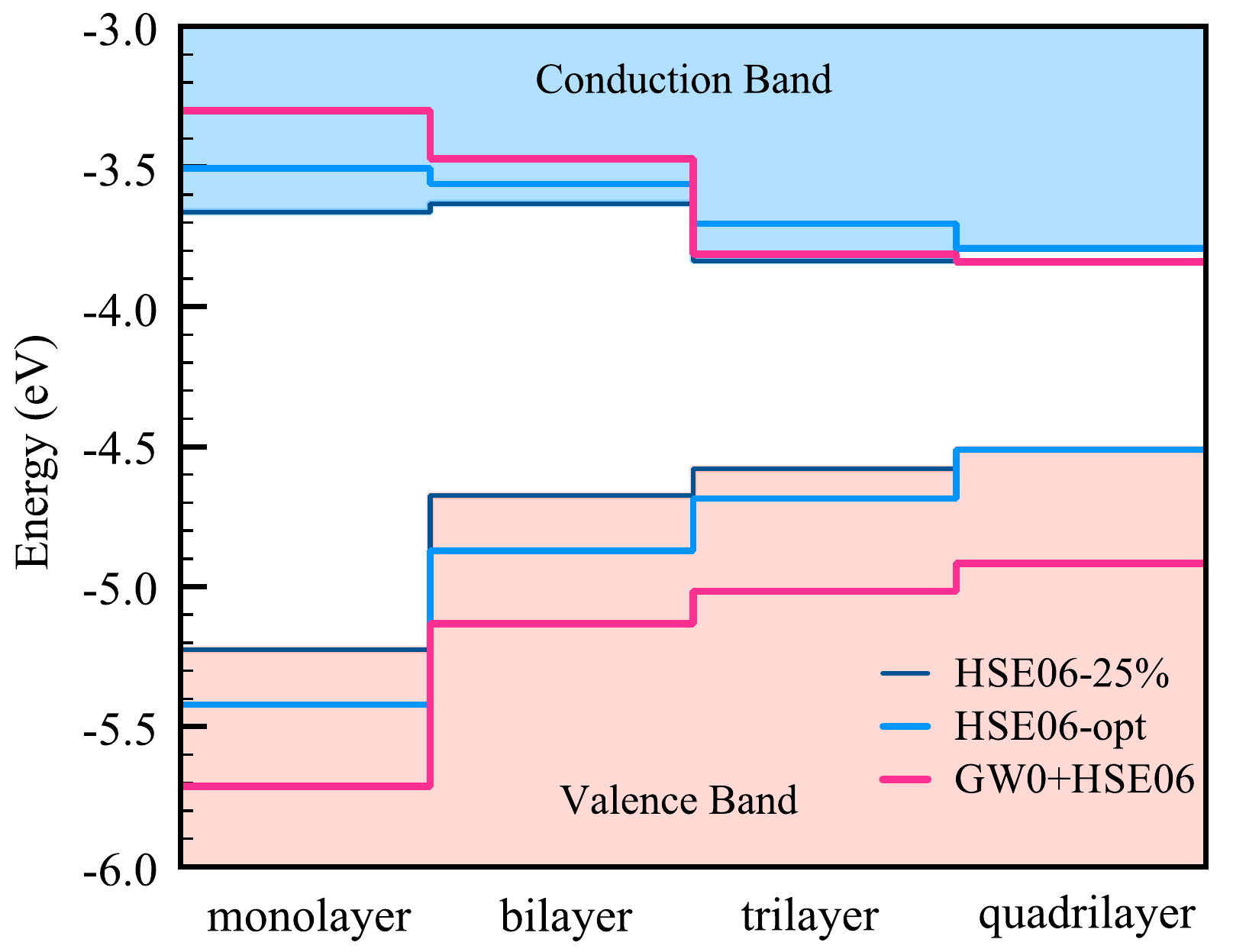}
\caption{\label{bandalign}(Color online) Band alignments for few-layer phosphorene. The vacuum level is taken as zero energy reference.}
\end{figure}

The calculated band alignments for few-layer phosphorene using difference approaches are shown in Fig. \ref{bandalign}. Although differing in magnitude, all approaches produce similar trends: (\emph{i}) with the increases in film thickness, the VBM and CBM of few-layer phosphorene move upward and downward respectively, as is the case in few-layer transition-metal dichalcogenides;\cite{Kang2013} (\emph{ii}) overall, the magnitude of band offset on the valence band is more significant than that on the conduction band. This implies that the transition levels of acceptors are more sensitively dependent on film thickness than those of donors.

\begin{figure}[htbp]
\centering
\includegraphics[scale=0.42]{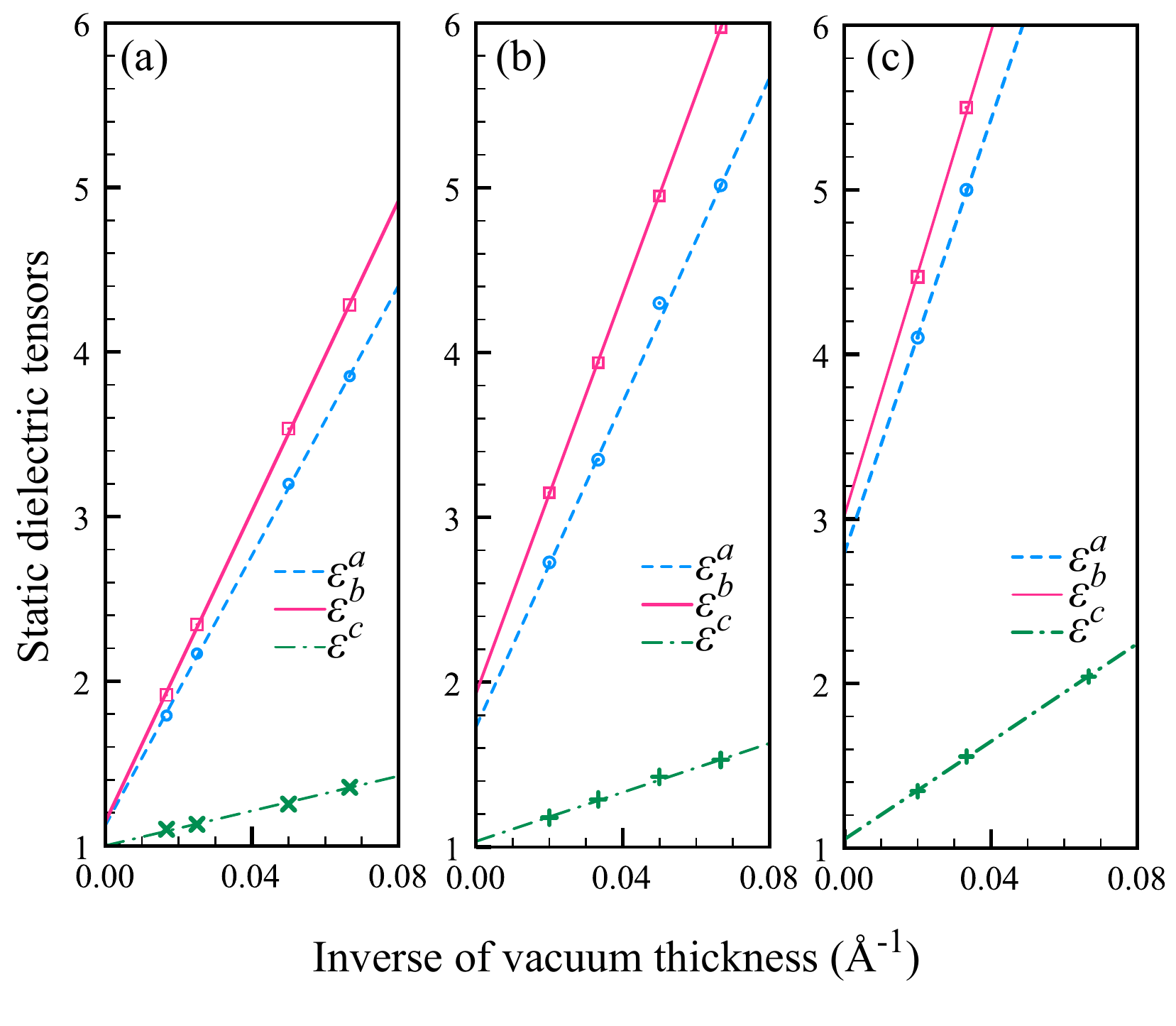}
\caption{\label{epsilon}(Color online) Static dielectric tensors $\epsilon$ as functions of the inverse of vacuum thickness for (a) monolayer, (b) bilayer and (c) quantrilayer phosphorene respectively.}
\end{figure}

To evaluate the formation energy of charged defects via Eq. (\ref{eq1}), we need to know the static dielectric tensor of few-layer phosphorene. With the periodic slab model, our calculated static dielectric constant tensor $\varepsilon$ demonstrate a linear dependence on the inverse of vacuum thickness (Fig. \ref{epsilon}). Obviously, the \emph{true} value of the static dielectric tensor is the one obtained in the limiting case of infinite vacuum. In effect, it can be extrapolated from the results for finite-size supercells with different vacuum thickness by scaling scheme. We list in Table \ref{dielectric} the calculated $\varepsilon$ of few-layer phosphorene parallel to \emph{a} ($\varepsilon^{a}$), \emph{b} ($\varepsilon^{b}$), and \emph{c} ($\varepsilon^{c}$) axes using HSE06. It is seen that the static dielectric tensor becomes larger for thicker phosphorene, due to enhanced screening effect. Additionally, the decrease in the band gap with increasing film thickness also contributes to this trend. The ionic contributions to the $\varepsilon$, on the other hand, are found to be rather small ($\leq$0.5). For the bulk system, our calculated $\varepsilon$ are noticeably different from Ref. \onlinecite{Asahina1984}, in which the frequency-dependent dielectric function calculations were performed using the local density approximation. 
Since the defective few-layer phosphorene systems have been modeled with supercells containing finite-size vacuum, the  $\varepsilon$ obtained from the corresponding pristine unit-cells have been adopted in calculating formation energies of defects.

\begin{table}[htbp]
\centering
\begin{ruledtabular}
\caption{\label{dielectric}  Static dielectric tensors $\varepsilon$ as a function of the inverse of vacuum thickness for (a) monolayer, (b) bilayer and (c) quatrilayer phosphorene.}
\begin{tabular}{c|ccc}
Systems & $\varepsilon^{a}$ & $\varepsilon^{b}$ & $\varepsilon^{c}$ \\
\hline
monolayer    &  1.12 &  1.15 & 1.01 \\
bilayer  &   1.72 & 1.93 & 1.03 \\
quadrilayer   & 2.79 & 3.02 & 1.05 \\
bulk  &   11.99 &  14.64 & 7.86 \\
bulk (Ref. \onlinecite{Asahina1984}) &   10.2 &  12.5 & 8.3 \\
\end{tabular}
\end{ruledtabular}
\end{table}

\subsection{Properties of native defects in few-layer phosphorene}
\begin{figure}[htbp]
\centering
\includegraphics[scale=0.34]{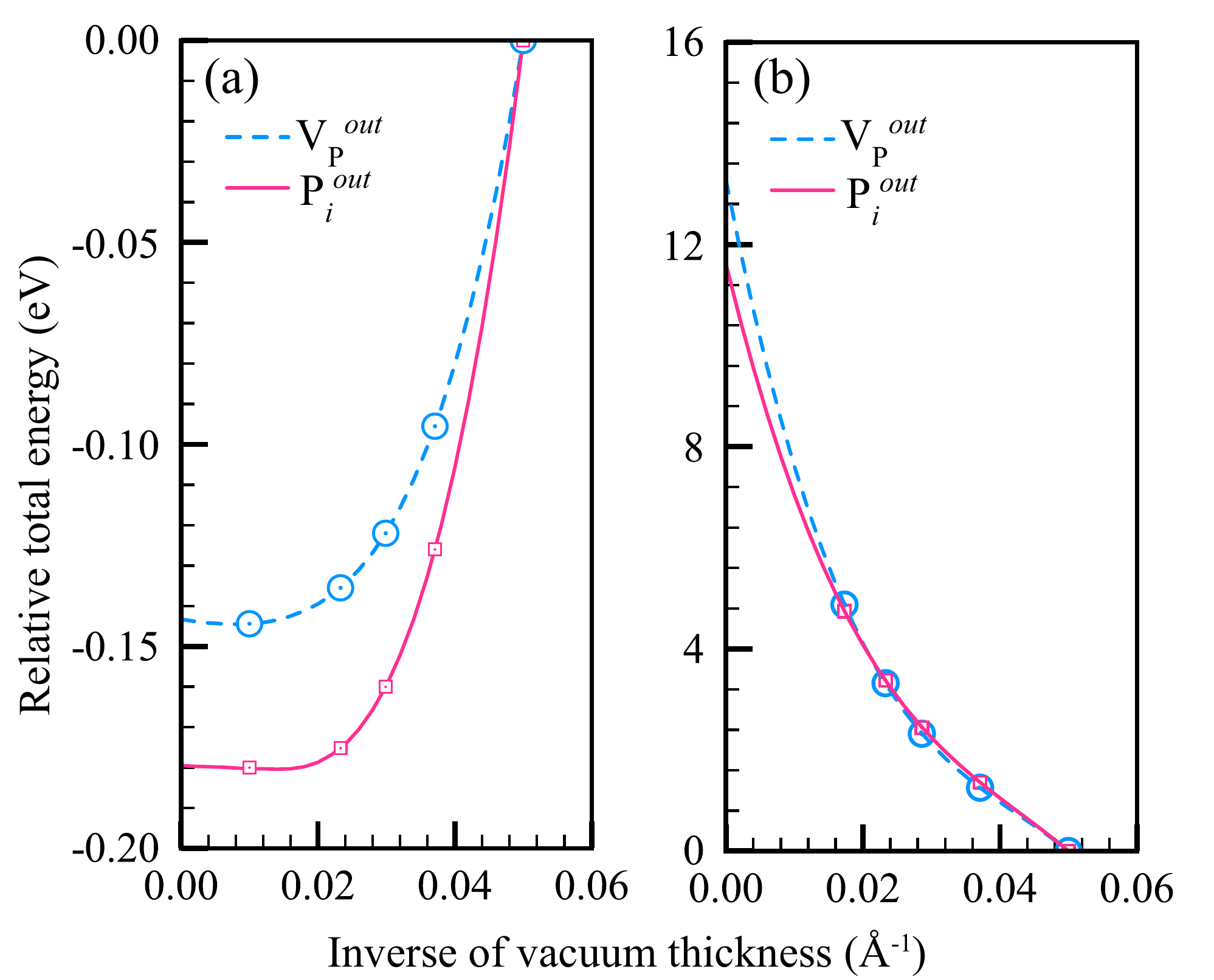}
\caption{\label{converge}(Color online) Total energies of phosphorene monolayer containing a vacancy, V$_\text{P}^{out}$, or a self-interstitial, P$_i^{out}$, in the charge states of (a) 1- or (b) 1+, as a function of the inverse of vacuum thickness. The total energies of the slabs with a vacuum thickness of 20 {\AA} are taken as zero.}
\end{figure}

In consideration of the electrostatic screening effect of vacuum slab along the \emph{c} direction is small (the dielectric constant of vacuum is equal to 1), we take monolayer phosphorene as an example to check the total energy convergence of charged-defect systems with respect to the vacuum thickness. Test calculations show that a vacuum thickness of 12 {\AA} can ensure the charge-neutral systems being well converged within 0.01 eV in total energies. This is not the case, however, for charged defects. Figure \ref{converge}(a) reveals that the numerical errors in the calculated total energies of monolayer phosphorene containing one V$_\text{P}^{out}$ or P$_i^{out}$ in 1- charge state are about 0.01 eV when a vacuum of 40 {\AA} was applied. However, for defects in 1+ charge state, a vacuum of 40 {\AA} is still far less than enough [Fig. \ref{converge}(b)]. Thus, the formation energies of positively and negatively charged native defects would be overestimated and underestimated in few-layer phosphorene when a typical 12 {\AA} vacuum was adopted without any corrections. These errors lead to unrealistic deeper transition levels for both acceptors and donors.

\begin{figure}[htbp]
\centering
\includegraphics[scale=0.43]{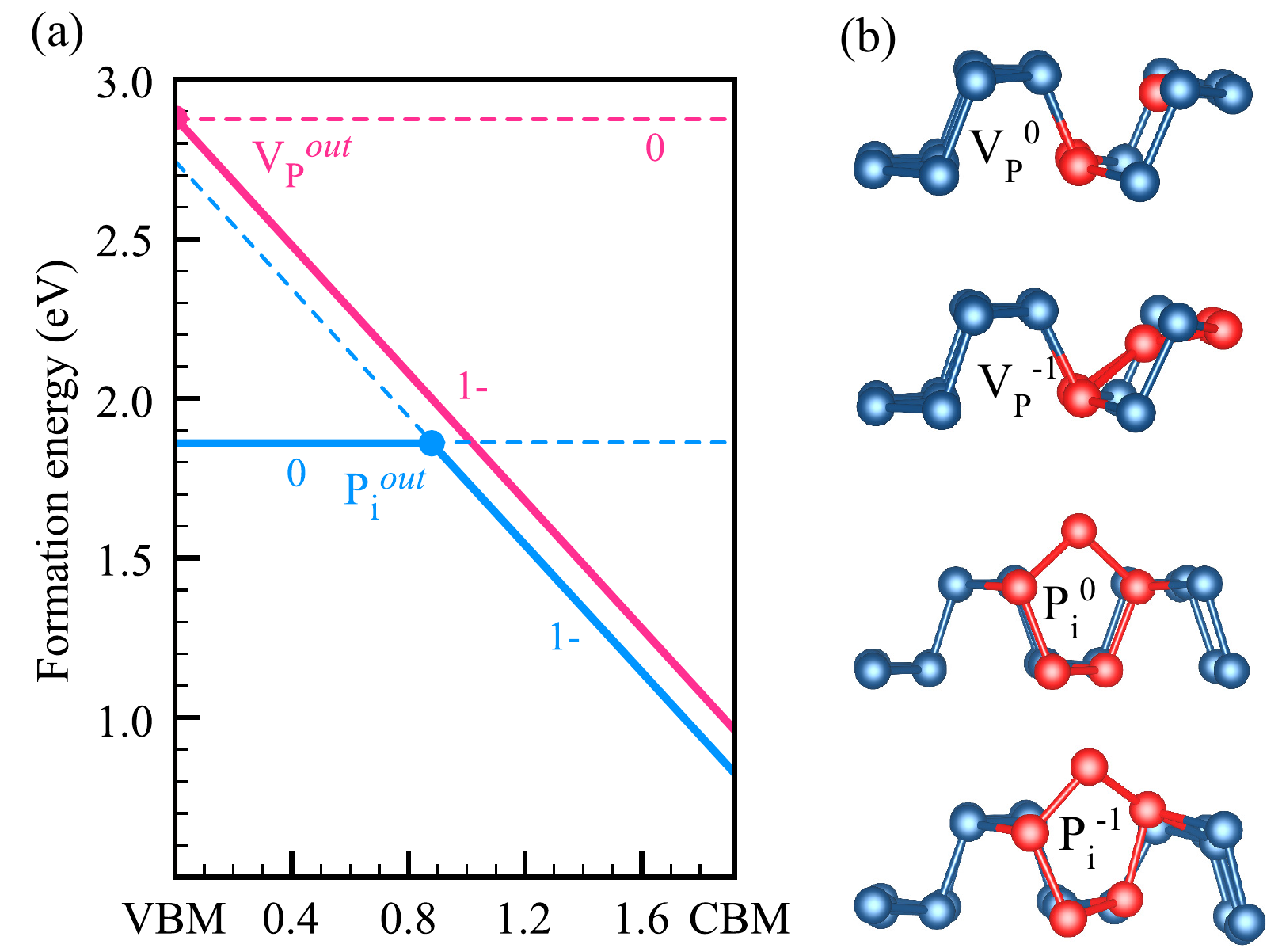}
\caption{\label{monolayer_D}(Color online) (a) Formation energy of V$_\text{P}$ and P$_i$ as a function of electron chemical potential $\mu_e$ in monolayer phosphorene. (b) Local structures of V$_\text{P}$ and P$_i$. The defect and its nearest-neighbors are colored differently.}
\end{figure}

The calculated formation energy of V$_\text{P}$ and P$_i$ in monolayer phosphorene as a function of electron chemical potential are plotted in Fig. \ref{monolayer_D}(a). The change of slope in the line for P$_i$ corresponds to the transition between charge states where thermodynamic transition takes place. We find that V$_\text{P}$ is stable in the charge state of 1- with respect to the neutral state for all values of \emph{E}$_\text{F}$ in the host band gap. This means that V$_\text{P}$ behaves as a shallow acceptor and could be one of the sources for \emph{p}-type conductivity observed experimentally.\cite{Liu2014} Because of the high formation energy (around 2.9 eV at \emph{E}$_\text{F}$=VBM), the negatively charged V$_\text{P}$ has a low concentration in phosphorene monolayer under equilibrium growth conditions, and thus might not be an efficient \emph{p}-type defect. Upon geometry optimization, the nearest neighbor of 1- charged V$_\text{P}$ on the top sublayer relaxes toward V$_\text{P}$ and bonds to its four neighbors with two different bond lengths of 2.41 {\AA} and 2.28 {\AA} respectively [Fig. \ref{monolayer_D}(b)]. It should be pointed that the donor ionization levels of V$_\text{P}$ or P$_i$ are unstable for all positions of \emph{E}$_\text{F}$ in the host band gaps of few-layer phosphorene, suggesting that both V$_\text{P}$ and P$_i$ are expected to be acceptors instead of donors.

A self-interstitial P atom, P$_i$, finds its stable position by bridging two host P atoms [see Fig. \ref{interstitial}(a)]. The formation energy of P$_i$ is about 1.0 eV lower than that of V$_\text{P}$ when \emph{E}$_\text{F}$ is near the VBM, suggesting that P$_i$ is the dominant native point defect under \emph{p}-type conditions. The (0/1-) acceptor level of P$_i$ is predicted to be 0.88 eV above the VBM, implying that P$_i$ is a deep acceptor. On the other hand, when the \emph{E}$_\text{F}$ is close to the host CBM, both V$_\text{P}$ and P$_i$ have much lowered formation energies and are energetically stable in the charge state of 1-. This means that they can serve as compensating centers in \emph{n}-type doping monolayer. In the neutral charge state, P$_i$ bonds to two host P atom with identical bond lengths of 2.14 {\AA}. A small asymmetry was observed in these two bonds (2.06 {\AA} versus 2.20 {\AA}), a local lattice distortion different from that around P$_i$ in 1- charge state. 

\begin{figure}[htbp]
\centering
\includegraphics[scale=0.4]{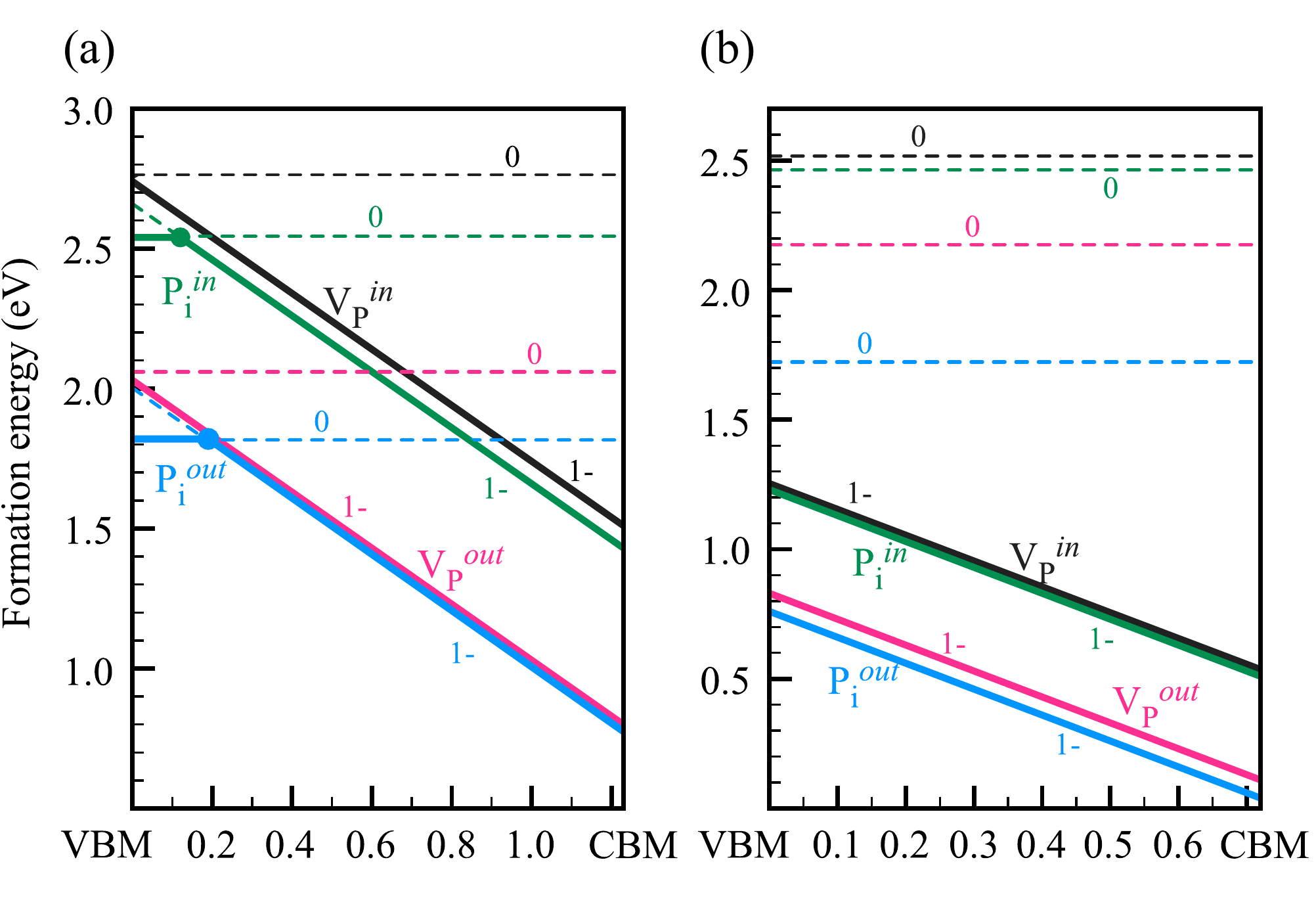}
\caption{\label{multilayer_D}(Color online) Formation energies of V$_\text{P}$ and P$_i$ in (a) bilayer and (b) quadrilayer phosphorene as a function of electron chemical potential.}
\end{figure}

In Figure \ref{multilayer_D}, we display the calculated formation energies of V$_\text{P}$ and P$_i$ in bilayer (panel a) and quadrilayer phosphorene (panel b) as a function of electron chemical potential. Our calculations show that both P$_i^{out}$ and V$_\text{P}^{out}$ are energetically more stable than P$_i^{in}$ and V$_\text{P}^{in}$ in both films, regardless of the charge states. The acceptor transition levels for V$_\text{P}^{out}$ and P$_i^{out}$ are -0.64 eV (not shown) and 0.19 eV with respect to the VBM, indicating that all possible native defects can contribute to the \emph{p}-type conductivity in bilayer. For quadrilayer phosphorene, both V$_\text{P}^{out}$ and P$_i^{out}$ are stable in the charge state of 1- for any \emph{E}$_\text{F}$ in the band gap. This trend is closely related to the upward shift of the band offset for VBM (see Fig. \ref{bandalign}). The calculated formation energies of all considered native defects decrease with the increase of film thickness. For examples, the calculated formation energies of the neutral V$_\text{P}^{out}$ and P$_i^{out}$ decrease from 2.88 and 1.86 eV in monolayer, to 2.67 and 1.82 eV in bilayer, and further to 2.18 and 1.73 eV in quanrilayer, with the layer-dependent effect being more significant on V$_\text{P}^{out}$ than P$_i^{out}$. Therefore, the formation energies of these acceptor-type defects in \emph{N}-layer phosphorene (\emph{N}>4) could be low enough when the \emph{E}$_\text{F}$ is near the CBM. As a result, self-compensation would be unavoidable in \emph{n}-type phosphorene. We expect that nonequilibrium growth techniques might be necessary to reduce the concentrations of native defects in preparation of \emph{n}-type phosphorene.

\begin{figure}[htbp]
\centering
\includegraphics[scale=0.34]{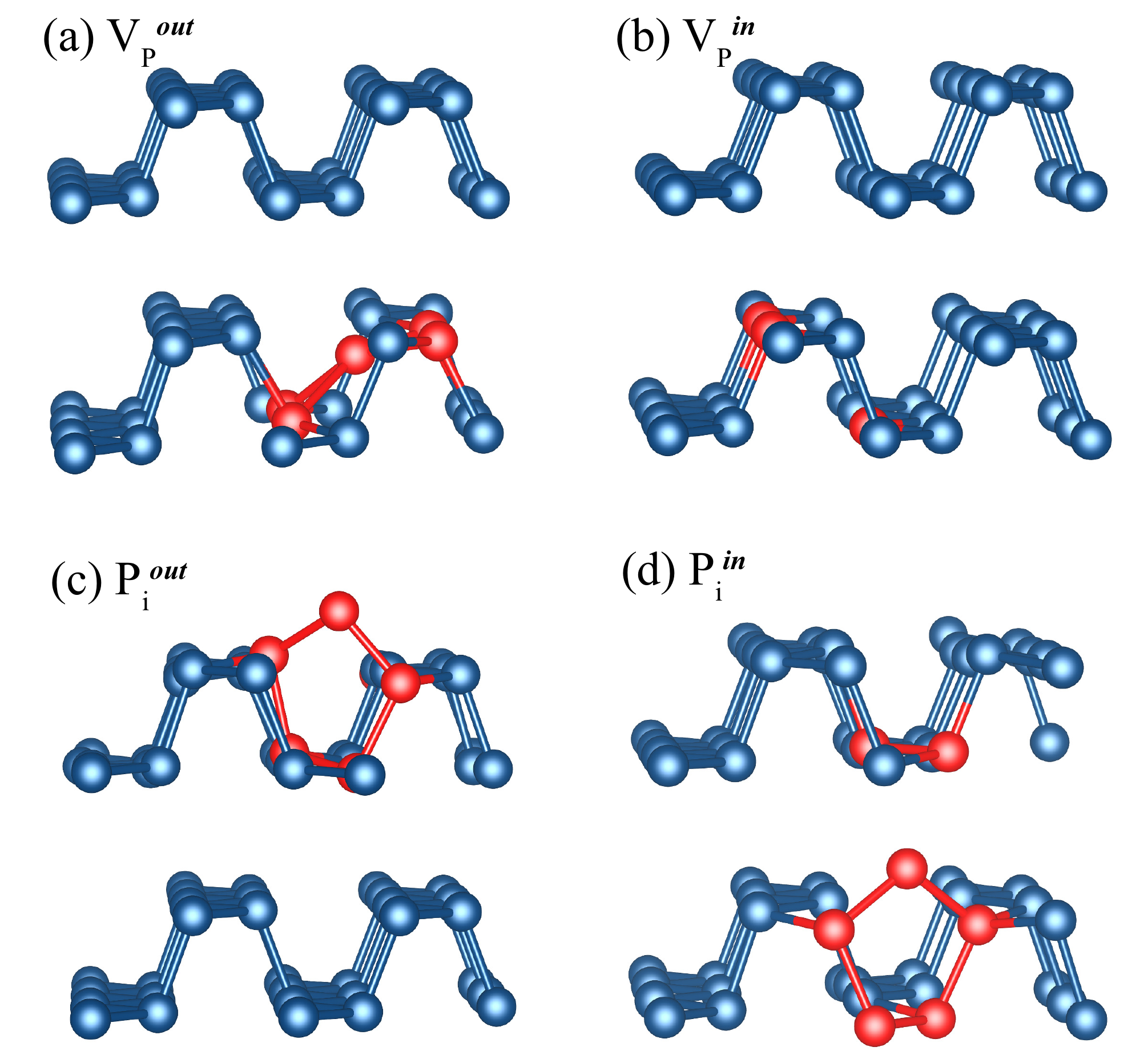}
\caption{\label{bilayer_struc}(Color online) Local structure of negatively charged (a) V$_\text{P}^{out}$, (b) V$_\text{P}^{in}$, (c) P$_i^{out}$ and P$_i^{in}$ in bilayer. The point defects and their nearest-neighbors are colored differently.}
\end{figure}

We plot in Fig. \ref{bilayer_struc} the local atomic arrangements around the negatively charged V$_\text{P}^{out}$, V$_\text{P}^{in}$, P$_i^{out}$ and P$_i^{in}$ in bilayer phosphorene. One can see that the relaxed local structure of V$_\text{P}^{out}$ is very similar to the case of monolayer (panel a). Unlike V$_\text{P}^{out}$, the neighboring P atoms of negatively charged V$_\text{P}^{in}$ undergo no significant distortion from their ideal lattice positions (panel b). This in turn leads to long ($\geq$3.1 {\AA}) and weak bonds between nearest-neighbors of V$_\text{P}^{in}$. The equilibrium local structure of negatively charged P$_i^{out}$ in bilayer is also similar to that in monolayer. As for P$_i^{in}$, the upper layer pushes the negatively charged P$_i^{in}$ to move downward, resulting in two identical bond lengths between P$_i^{in}$ and its two nearest-neighbors (2.14 {\AA}). Meanwhile, the nearest-neighbors on the upper layer relax symmetrically away from P$_i^{in}$, as illustrated in panel (d).

\begin{figure}[htbp]
\centering
\includegraphics[scale=0.5]{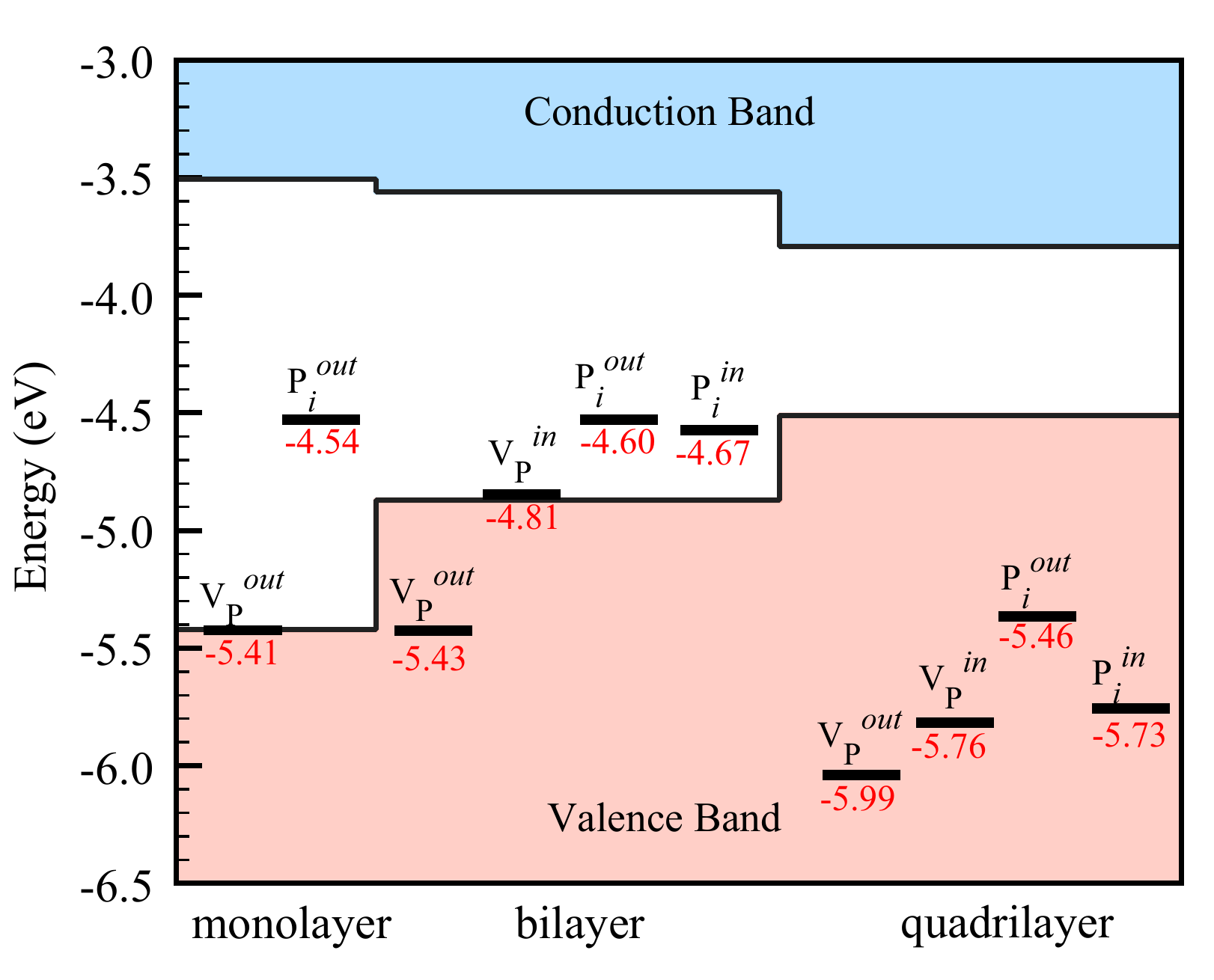}
\caption{\label{defectalign}(Color online) Transition levels of V$_\text{P}$ and P$_i$ in few-layer phosphorene, referenced to the vacuum level.}
\end{figure}

To gain deeper insight into the origin of the conductive characteristics in few-layer phosphorene, we display in Fig. \ref{defectalign} the transition levels of native point defects with respect to the vacuum level. One can see that the transition levels of V$_\text{P}$ and P$_i$ generally decrease with increasing film thickness. This means that the magnitudes of formation energies of negatively charged defects decrease more rapidly than those for the neutral ones when going from monolayer to quadrilayer. This results in the shift of the acceptor transition levels of V$_\text{P}$ and P$_i$ toward lower energies. Combined with the band offset effects for the host VBM, this shift is also responsible for the observed shallower acceptor levels of V$_\text{P}$ and P$_i$ in thicker films.

\begin{figure}[htbp]
\centering
\includegraphics[scale=0.4]{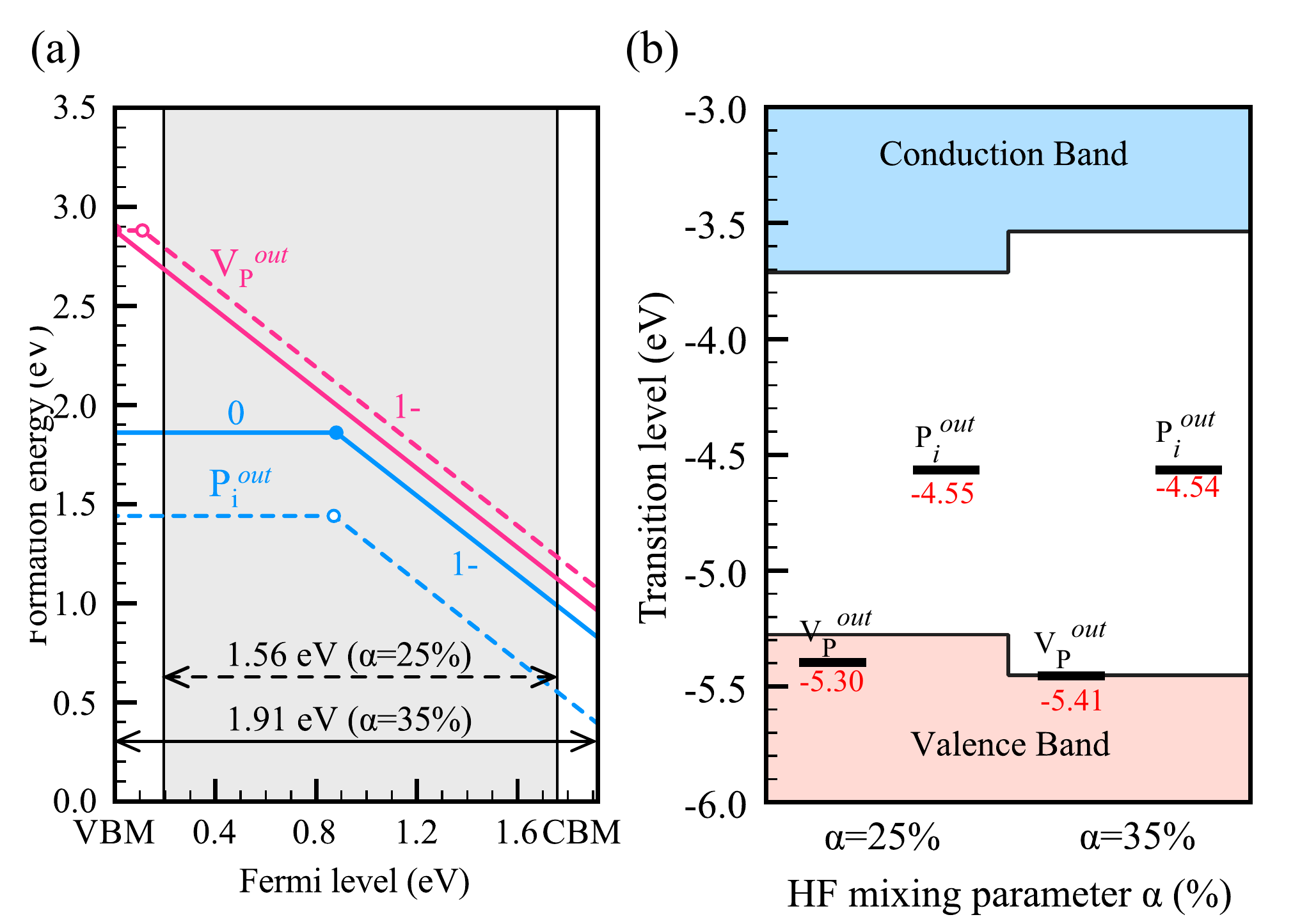}
\caption{\label{alpha}(Color online) Formation energies of V$_\text{P}$ and P$_i$ as functions of electron chemical potential in monolayer phosphorene given by HSE06-$\alpha$ method. The solid and dashed lines represent $\alpha$=35\% and $\alpha$=25\% results. The gray region represents the HSE06-25\% band gap. (b) Transition energy referenced to the vacuum level.}
\end{figure}

We note that three different HF mixing parameter $\alpha$ (25\%, 30\% and 35\%) were adopted for monolayer, bilayer and quadrilayer phosphorene. We now take V$_\text{P}^{out}$ and P$_i^{out}$ as examples to investigate the impact of $\alpha$ on their stability and conductivity. We present in panel (a) of Fig. \ref{alpha} the comparison of HSE06-25\% and HSE06-35\% in respect of the formation energy of V$_\text{P}$ and P$_i$ as a function of electron chemical potential in monolayer phosphorene. A deviation of around 0.4 eV is observed for the formation energy of P$_i^{out}$; while the change in transition levels of V$_\text{P}^{out}$ and P$_i^{out}$, shown in panel (b), is within 0.1 eV when $\alpha$ goes from 35\% to 25\%. This suggests that $\alpha$ has insignificant effects on the transition levels of V$_\text{P}^{out}$ and P$_i^{out}$. The rigid shifts of the host VBM are primarily responsible for the shallower transition levels which are calculated by using HSE06-25\% approach. Furthermore, one can conclude that P$_i$ still acts as a deep acceptor if the monolayer phosphorene  has a band gap value of 1.56 eV, based on the HSE06-25\% calculated results. 
We  expect it to hold true for thicker phosphorene. Similar results were also found for the native defects in GaInO$_3$.\cite{Wang2015}

\section{summary}
In conclusion, we have investigated the structural and electronic properties of native point defect in few-layer phosphorene using first-principles calculations based on hybrid density functional theory including vdW correction. Our calculations show that both vacancy and self-interstitial P defects exhibit acceptor-like behavior and their formation energies and transition levels decrease with increasing film thickness. The same trend is also observed in the host band gap. These trends can be explained by the band offsets for few-layer phosphorene. Specifically, we find that the valence band maximum and conduction band minimum systematically shift upward and downward in reference to the vacuum level with the increases of film thickness. As a result, both vacancies and self-interstitials become shallow acceptors in few-layer phosphorene and can acount for the sources of \emph{p}-type conductivities observed in experiments. On the other hand, these native acceptors could have non-negligible concentrations and thus act as compensating centers in \emph{n}-type phosphorene.

\begin{acknowledgments}
We thank Y. Kumagai for valuable discussions. V. Wang acknowledges the support of the Natural Science Foundation of Shaanxi Province, China (Grant no. 2013JQ1021). Y. Kawazoe is thankful to the Russian Megagrant Project No.14.B25.31.0030 ``New energy technologies and energy carriers''
for supporting the present research. The calculations were performed on the HITACHI SR16000 supercomputer at the Institute for Materials Research of Tohoku University, Japan.
\end{acknowledgments}

\nocite{*}
\bibliographystyle{aipnum4-1}
%
\end{document}